\documentclass[twoside,english]{iopart}
\usepackage[T1]{fontenc}
\usepackage[latin9]{inputenc}
\usepackage{geometry}
\usepackage{amsbsy}
\usepackage{amstext}
\usepackage{graphicx}
\usepackage{esint}

\makeatletter
\usepackage{iopams}
\usepackage{setstack}

\usepackage[numbers, sort&compress]{natbib}

\newcommand{\eqref}[1]{(\ref{#1})}

\makeatother

\usepackage{babel}
\begin{document}

\title{Nonlinear Schr{\"o}dinger equation from generalized exact uncertainty principle }

\author{{\Large{}\L ukasz Rudnicki}}

\address{{\large{}Institute for Theoretical Physics, University of Cologne, Z{\"u}lpicher Stra{\ss}e 77, D-50937, Cologne, Germany}}

\address{{\large{}Center for Theoretical Physics, Polish Academy of Sciences, Aleja Lotnik{\'o}w 32/46, PL-02-668 Warsaw, Poland}}

\ead{{\large{}rudnicki@cft.edu.pl}}
\begin{abstract}
Inspired by the generalized uncertainty principle (GUP), which adds gravitational effects to the standard description of quantum uncertainty, we extend the exact uncertainty principle (EUP) approach by Hall and Reginatto {[}J. Phys. A: Math. Gen. (2002) \textbf{35} 3289{]}, and obtain a (quasi)nonlinear Schr{\"o}dinger equation. This quantum evolution equation of unusual form, enjoys several desired properties like separation of non-interacting subsystems or plane-wave solutions for free particles. Starting with the harmonic oscillator example, we show that every solution of this equation respects the gravitationally-induced minimal position uncertainty proportional to the Planck length. Quite surprisingly, our result successfully merges the core of classical physics with non-relativistic quantum mechanics in its extremal form. We predict that the commonly accepted phenomenon, namely a modification of a free-particle dispersion relation due to quantum gravity might not occur in reality. 
\end{abstract}
\maketitle

\section{Introduction}

Even though, the famous Schr{\"o}dinger equation does not provide the most general description of quantum systems (for instance, only approximates the Dirac equation) it remains useful while studying fundamental aspects of quantum mechanics. A prominent example aspect discussed in this contribution is the \emph{linearity of Quantum Mechanics} (superposition principle) which, if valid universally, shall also apply to the Schr{\"o}dinger equation in its pure form. We thus do not consider here the Gross-Pitaevski (describing Bose-Einstein condensate) or Schr{\"o}dinger-Newton (self-gravity effects in Newtonian approximation) extensions of the standard non-relativistic quantum dynamics, but focus on a nonlinearity\emph{ per se}, possibly present in the genuine Schr{\"o}dinger equation.

In the literature (see a rare example \cite{NotBEC} not related to Bose-Einstein condensate) on the field one can mainly find the discussion of the quadratic (in the wavefunction $\psi\left(x,t\right)$; by $x$ we denote an $n$-dimensional position vector) nonlinearity included as an additional term proportional to the probability density $\rho=\left|\psi\right|^{2}$. Most of formal research focuses on the mathematical aspects of this basic nonlinearity (integrability, blow up, etc. \cite{Blow,SmallReview,numerics}), while more interdisciplinary approaches apply the resulting wave equation beyond quantum physics \cite{water,water2}. In a more general scenario, the nonlinearity is introduced by a function of $\rho$, commonly (but not always) being equal to $0$ for $\rho=0$ (see \cite{SmallReview} for few interesting examples). Note that in order to mimic the structure of the linear equation, every discussed correction is always multiplied by the wavefunction, so that all nonlinear contributions play the role of state (or only density) dependent potentials.

As pointed out by Bialynicki-Birula \cite{BirulaBraz}, nonlinear modifications of the above type (except one):
\begin{enumerate}
\item Introduce an extra interaction between separable subsystems,
\item Spoil the standard normalization procedure for stationary states.
\end{enumerate}
While the first issue can sometimes be accepted as an emanation of a possibly unavoidable link between subsystems one intends to separate, the second issue diminishes the beauty of mathematical analogy between quantum states and projective rays of the Hilbert space. In the comprehensive discussion devoted to formally reasonable nonlinearities \cite{Weinberg}, the second argument also referred to as lack of homogeneity, led to the conclusion that all inhomogeneous proposals are actually not of physical relevance. On the other hand some homogeneous nonlinearities, like $(\left|\boldsymbol{\nabla}\psi\right|/\left|\psi\right|)^{2}$ discussed by Kibble \cite{Kibble}, bring on board the third issue, namely they\begin{enumerate}\setcounter{enumi}{2}\item Violate Galilean invariance. \end{enumerate}

The single form of nonlinearity, depending only on the density and free from the above limitations, is given by the logarithmic term $-b\log\rho$. It was long ago shown \cite{BirulaMech,IBBmore,IBBmore2} that the parameter $b$ if different from zero must be very small, at least $b<4\times10^{-10}\textrm{eV}$. Experimental tests \cite{Zeilinger} put a more rigorous limitation $b<3.3\times10^{-15}\textrm{eV}$ (for a short summary of other obtained estimations see \cite{BirulaBraz}). It is worth noticing that the logarithmic correction was later derived on the ground of the stochastic equation \cite{stochast}.

It is rather obvious that possible nonlinearities of the Schr{\"o}dinger equation (if any) must in normal conditions contribute in a negligible manner. On the other hand, in extraordinary situations (eg. very high energies), when the conceivable nonlinearities could play any noticeable role, the Schr{\"o}dinger equation will likely be an insufficient approximation. Nevertheless, the question whether the pure Schr{\"o}dinger equation contains any nonlinearities (even of extremely low contribution) remains of fundamental interest, as it challenges the superposition principle. Moreover, nonlinear Schr{\"o}dinger dynamics can be interesting from the perspective of down-to-earth problems such as quantum state discrimination \cite{discrimination}.

In the current paper, instead of \emph{ad hoc} proposing a new form of a suitable nonlinearity, we follow and slightly generalize the approach of \emph{exact uncertainty principle} (EUP) by Hall and Reginatto \cite{Hall,Hall2,Hall3,Hall4,Reginatto}. In their seminal paper \cite{Hall}, the authors have postulated a fundamental scaling relation between nonclassical momentum fluctuations and uncertainty in position (please refer to Section \ref{sec:Quantum-dynamics-from} for details), which is a prerequisite for the Heisenberg uncertainty relation (HUR), $\Delta x\Delta p\geq\hbar/2$, involving position and momentum standard deviations. 

Our proposed generalization is driven by a reverted reasoning. In the regime relevant for potential nonlinearities of quantum dynamics, also the HUR might likely require a modification. A model example of this theoretically predicted phenomenon is the family of generalized uncertainty principles (GUP) with its most basic member given by \cite{GUP1,GUP2,GUPPRD,GUPreview} 
\begin{equation}
\Delta x\Delta p\geq\frac{\hbar}{2}\left[1+\beta\left(\Delta p\right)^{2}\right].\label{GUP}
\end{equation}
The right hand side of the above inequality contains corrections depending on the momentum uncertainty, introduced by the parameter $\beta=\beta_{0}l_{p}^{2}/\hbar^{2}\equiv\beta_{0}G/(\hbar c^{3})$ encoding gravitational effects. $l_{p}=1.62\times10^{-35}\textrm{m}$ denotes the Planck length, $G$ is the Newton constant while $\beta_{0}$ is a numerical parameter (likely of order of unity) depending on the approach towards quantization of gravity. In this contribution I however would not like to discuss more and rely on the theoretical foundations of the GUP (interested readers shall consult the comprehensive review \cite{GUPreview}), since violation of the Lorentz invariance present in the modified energy-momentum dispersion relations or in doubly special relativity does not belong to the set of commonly accepted laws of nature. I find it sufficient to expect, that due to various possible physical reasons such as potential existence of maximal proper acceleration \cite{properA} (note that this preserves Lorentz invariance \cite{properA2}), the equation (\ref{GUP}) or a similar one replaces the HUR \cite{GUPpA}. Let me only observe that the mathematical structure of (\ref{GUP}) and its counterparts, leads by a straightforward optimization to the minimal measurable length \cite{GUPPRD}, or in other cases to existence of minimal measurable momentum, minimal time interval or maximal measurable energy \cite{Tawfik1}. Such constrains, when considered, influence various physical predictions such as description of black holes \cite{Tawfik2} or Newton's laws of gravity \cite{Tawfik3}.

As we will see later, since (\ref{GUP}) is sharper than the HUR, a suitable modification of the EUP is a natural path to follow. The exact uncertainty principle, adjusted to a sharper uncertainty relation (like the GUP, but not only), necessarily brings (quasi)nonlinear corrections to the Schr{\"o}dinger equation, which in a quite general scenario becomes ($V(x)$ is a potential):
\begin{equation}
i\hbar\frac{\partial\psi}{\partial t}=\left(-\frac{\hbar^{2}}{2m}\nabla^{2}+V-\frac{\hbar^{2}}{2m}\sum_{l=1}^{n}W\left(CF_{l}\left[\rho\right]\right)\frac{1}{\left|\psi\right|}\frac{\partial^{2}\left|\psi\right|}{\partial x_{l}^{2}}\right)\psi.\label{NLSchr}
\end{equation}
$W\left(\cdot\right)$ is a dimensionless function associated with the uncertainty relation considered (please see Eqs. \ref{SharperHUR} and \ref{W} in Section \ref{sec:Generalized-Exact-Uncertainty}) and expected to assume values significantly smaller than $1$. By construction, it has a property $W(0)=0$. The functional
\begin{equation}
F_{l}\left[\rho\right]=\int d^{n}x\,\frac{1}{\rho}\left(\frac{\partial\rho}{\partial x_{l}}\right)^{2},\label{Fisher}
\end{equation}
is the Fisher information with respect to the variable $x_{l}$, while $C=\hbar^{2}/4$.

I call the above equation quasi-nonlinear because of the specific forms of nonlinearity appearing in its last term. First of all, the dependence on $\rho$ via the function $W$ is hidden inside the space integrals present in $F_{l}\left[\rho\right]$. This means, that while looking for stationary states 
\begin{equation}
\psi\left(x,t\right)=e^{-iEt/\hbar}\psi\left(x\right),\label{stationary}
\end{equation}
for which $\rho$ is time-independent, one can treat $W_{l}\equiv W\left(CF_{l}\left[\rho\right]\right)$ as being a collection of fixed parameters, and attempt to find the solution in the form $\psi\left(x;W_{1},\ldots,W_{n}\right)$. Then, one is left with a system of $n$ algebraic equations for the $W_{l}$ parameters, given by the consistency conditions utilizing the definition (\ref{Fisher}) and the function $W$, namely:
\begin{equation}
W_{l}=W\left(CF_{l}[\left|\psi\left(x;W_{1},\ldots,W_{n}\right)\right|^{2}]\right).\label{consistency}
\end{equation}
Moreover, even though $\left|\psi\right|=\sqrt{\rho}$, in the common case when $\psi\left(x\right)$ from (\ref{stationary}) is real, the derivative term reduces to the standard (linear) form with the second derivative of $\psi(x)$. In this scenario we simply recover the usual, time-independent Schr{\"o}dinger equation with each component of the Laplacian multiplied by the constant $1+W_{l}$ (effective mass). We thus obtain an almost linear situation with the single exception, that at the end the consistency equations (\ref{consistency}) shall be solved.

Remarkably, the nonlinear Eq. \ref{NLSchr} enjoys all three desired properties, namely separability of non-interacting systems, the norm invariance (homogeneity) and Galilean invariance. If one substitutes: 
\begin{equation}
\psi\left(x,t\right)=\prod_{l=1}^{n}\psi_{l}\left(x_{l},t\right),\label{separable}
\end{equation}
then $F_{l}\left[\rho\right]=F_{l}\left[\rho_{l}\right]$, with $\rho_{l}=\left|\psi_{l}\right|^{2}$, and $\left|\psi\right|^{-1}\partial_{l}^{2}\left|\psi\right|$ depends only on $\left|\psi_{l}\right|$. The full, time-dependent dynamics of the state (\ref{separable}) becomes separable, provided the potential contains no interaction between subsystems. Any Galilean transformation shifts the time derivative and adds a position-dependent phase to the wavefunction. The Fisher information is invariant as the functional of the density, while the second-derivative terms depend on $\left|\psi\right|$ (again the phase does not contribute). In relation to the norm invariance for the stationary states (\ref{stationary}), if $\psi\left(x;W_{1},\ldots,W_{n}\right)$ is a solution valid before the application of the consistency conditions (\ref{consistency}), then also $A\psi\left(x;W_{1},\ldots,W_{n}\right)$ is the proper solution in the same case. One can normalize this solution in a standard way, so that the constant $A$ acquires a dependence on $W_{l}$. Actually, the algebraic conditions (\ref{consistency}) can be written down and possibly solved only after the normalization procedure.

Obviously the true dynamics spoils the norm invariance letting the nonlinearity have a genuine character. Since the superposition principle does not hold, the stationary states do not provide a frame to study the time evolution. Note also, what is to be expected, that Eq. \ref{NLSchr} reduces to the pure Schr{\"o}dinger equation if the solution is a plane wave (in this case $F_{l}[\rho]=0$).

In the next section we shall bring on board the basic results concerning the approach based on the exact uncertainty principle merged with classical dynamics. In Section \ref{sec:Generalized-Exact-Uncertainty} we generalize the EUP in order to handle the modified forms of the HUR and derive the corresponding quantum dynamics. In the last section we discuss the Gaussian solution of Eq. \ref{NLSchr} for the 1D harmonic potential, with special emphasis on the GUP case.

\section{Quantum dynamics from exact uncertainty principle\label{sec:Quantum-dynamics-from} }

We begin this short review of the EUP formalism by a list of all relevant quantities and concepts. Except an expanded discussion (see Sec. \ref{sec:Fluct}) of technical assumptions relevant for the momentum fluctuations, followed by their mild modifications, the material presented below is a summary of sections 2 and 3 from \cite{Hall}. The starting point is the classical dynamics described in terms of the Hamilton's principal function $S\left(x,t\right)$ and the probability density $P\left(x,t\right)$. The function $S$ evolves according to the Hamilton-Jacobi equation 
\begin{equation}
\frac{\partial S}{\partial t}+\frac{1}{2m}\boldsymbol{\nabla}S\cdot\boldsymbol{\nabla}S+V=0,
\end{equation}
and its coordinate derivatives (components of $\boldsymbol{\nabla}S$) give the classical momentum $\boldsymbol{p}_{\textrm{cl}}$. The probability density satisfies the continuity equation
\begin{equation}
\frac{\partial P}{\partial t}+\boldsymbol{\nabla}\cdot\left(P\frac{1}{m}\boldsymbol{\nabla}S\right)=0,\label{continuity}
\end{equation}
since $m^{-1}\boldsymbol{\nabla}S$ is a velocity field. Both equations can be derived by virtue of a variational principle applied to the classical action 
\begin{equation}
\mathcal{A}_{C}=\int dtd^{n}x\, P\left[\frac{\partial S}{\partial t}+\frac{1}{2m}\boldsymbol{\nabla}S\cdot\boldsymbol{\nabla}S+V\right].
\end{equation}

The quantum dynamics (standard Schr{\"o}dinger equation) emerges from the classical action modified by a term brought by the EUP. To make a long story short (for all other details see \cite{Hall}) the quantum momentum is described as a fluctuating classical momentum: $\boldsymbol{p}=\boldsymbol{p}_{\textrm{cl}}+\boldsymbol{N}$, or componentwise ($l=1,\ldots,n$)
\begin{equation}
p_{l}=\frac{\partial S}{\partial x_{l}}+N_{l}.\label{momentum}
\end{equation}
The fluctuation term $\boldsymbol{N}$ satisfies certain randomness assumptions (see Sec. \ref{sec:Fluct}) and, as a result, modifies the action to the quantumly corrected form
\begin{equation}
\mathcal{A}_{Q}=\mathcal{A}_{C}+\frac{1}{2m}\sum_{l=1}^{n}\int dt\left(\Delta N_{l}\right)^{2},\label{Aq}
\end{equation}
with each $\Delta N_{l}$ being the root-mean-square fluctuation in $l$th direction, averaged over the $n$-dimensional coordinate space.

\subsection{Properties of momentum fluctuations\label{sec:Fluct} }

The extra ingredient of the above formalism, namely the momentum fluctuations $\boldsymbol{N}$, shall be subject to physically motivated restrictions. To this end, the authors of \cite{Hall} distinguished two types of averages. Given a random, position-dependent quantity $\Phi\left(x\right)$ one can either only average out the fluctuations at a given point --- obtaining the position-dependent field $\bar{\Phi}\left(x\right)$ --- or go one step further and average the resulting field over the position space:
\begin{equation}
\left\langle \Phi\right\rangle =\int d^{n}x\, P(x)\bar{\Phi}\left(x\right).
\end{equation}
Reasonably behaving fluctuations shall on average vanish in every point \cite{Hall}, that is $\bar{\boldsymbol{N}}=0$. The authors of \cite{Hall} observed that for the discussed purpose it is sufficient to impose a couple of substantially weaker conditions (see Eq. 5 from \cite{Hall})
\begin{equation}
\left\langle \boldsymbol{N}\right\rangle =0,\qquad\left\langle \boldsymbol{\nabla}S\cdot\boldsymbol{N}\right\rangle \equiv\left\langle \boldsymbol{p}_{\textrm{cl}}\cdot\boldsymbol{N}\right\rangle =0.\label{unbiasedness}
\end{equation}
These two conditions also have an appealing meaning, namely the first one says that fluctuations disappear on the total average $\left\langle \cdot\right\rangle $ while the second one imposes \emph{unbiasedness} between the fluctuations in question and the classical momentum. 

Eq. \ref{Aq} appears as a consequence of the replacement 
\begin{equation}
\int d^{n}x\, P(x)\boldsymbol{\nabla}S\cdot\boldsymbol{\nabla}S\equiv\left\langle \boldsymbol{p}_{\textrm{cl}}\cdot\boldsymbol{p}_{\textrm{cl}}\right\rangle \mapsto\left\langle \boldsymbol{p}\cdot\boldsymbol{p}\right\rangle =\left\langle \left(\boldsymbol{p}_{\textrm{cl}}+\boldsymbol{N}\right)\cdot\left(\boldsymbol{p}_{\textrm{cl}}+\boldsymbol{N}\right)\right\rangle ,\label{replacement}
\end{equation}
and 
\begin{equation}
\left\langle \left(\boldsymbol{p}_{\textrm{cl}}+\boldsymbol{N}\right)\cdot\left(\boldsymbol{p}_{\textrm{cl}}+\boldsymbol{N}\right)\right\rangle =\left\langle \boldsymbol{p}_{\textrm{cl}}\cdot\boldsymbol{p}_{\textrm{cl}}\right\rangle +2\left\langle \boldsymbol{p}_{\textrm{cl}}\cdot\boldsymbol{N}\right\rangle +\left\langle \boldsymbol{N}\cdot\boldsymbol{N}\right\rangle ,
\end{equation}
where $\left\langle \boldsymbol{N}\cdot\boldsymbol{N}\right\rangle =\left(\Delta N\right)^{2}$ are the \emph{total fluctuations}:
\begin{equation}
\Delta N=\sqrt{\sum_{l=1}^{n}\left(\Delta N_{l}\right)^{2}}.\label{deltaN}
\end{equation}
It is important to understand that mathematically (to derive Eq. \ref{Aq}), one needs to eliminate the mixed term $\left\langle \boldsymbol{p}_{\textrm{cl}}\cdot\boldsymbol{N}\right\rangle $. This goal, however, can be achieved in many ways. One can impose the stronger condition $\bar{\boldsymbol{N}}=0$ which works because $\boldsymbol{p}_{\textrm{cl}}$ is free from fluctuations, or by resorting to the unbiasedness assumption which literally says that the mixed term in question vanishes. Both ways are physically reasonable, and can thus be used interchangeably. For instance, in \cite{Hall3} only the stronger condition is utilized.

Staying on the purely mathematical ground, we could find more possibilities leading to the same result. For example, one can consider the momentum being a complex field, such that the classical momentum is real while the momentum fluctuations are purely imaginary. In the replacement (\ref{replacement}) one would then need to use a complex conjugate field as well, $\left\langle \boldsymbol{p}_{\textrm{cl}}\cdot\boldsymbol{p}_{\textrm{cl}}\right\rangle \mapsto\left\langle \boldsymbol{p}^{*}\cdot\boldsymbol{p}\right\rangle $. In this way, the ``desired'' result follows for any (imaginary) $\boldsymbol{N}$ and without randomness assumptions. One shall argue that introducing the complex numbers would spoil the classical flavor of the whole derivation. On the other hand Quantum Mechanics relies on the complex wave vectors and functions, while the momentum and other observables are Hermitian (though mildly complex) operators.

In the current contribution we shall not further explore the path of complex replacement, but we will rely on the randomness assumptions presented above. We would like to observe, however, that the single unbiasedness assumption is not enough from the perspective of the HUR. One could easily find that
\begin{equation}
\left(\Delta p\right)^{2}=\left\langle \boldsymbol{p}\cdot\boldsymbol{p}\right\rangle -\left\langle \boldsymbol{p}\right\rangle \cdot\left\langle \boldsymbol{p}\right\rangle =\left\langle \boldsymbol{p}_{\textrm{cl}}\cdot\boldsymbol{p}_{\textrm{cl}}\right\rangle +2\left\langle \boldsymbol{p}_{\textrm{cl}}\cdot\boldsymbol{N}\right\rangle +\left\langle \boldsymbol{N}\cdot\boldsymbol{N}\right\rangle -\left\langle \boldsymbol{p}_{\textrm{cl}}\right\rangle \cdot\left\langle \boldsymbol{p}_{\textrm{cl}}\right\rangle ,
\end{equation}
where $\left\langle \boldsymbol{p}\right\rangle =\left\langle \boldsymbol{p}_{\textrm{cl}}\right\rangle $ as $\left\langle \boldsymbol{N}\right\rangle =0$. Due to the unbiasedness assumption, we obtain $\Delta p\geq\Delta N$, which as discussed in the next subsection is the prerequisite to the HUR. 

The single unbiasedness assumption it too weak to render the componentwise relations $\Delta p_{l}\geq\Delta N_{l}$ satisfied independently for all $l\in\left\{ 1,\ldots,n\right\} $. Since these inequalities are necessary to assure the validity of the individual HURs (for any position-momentum couple) we shall strengthen the unbiasedness condition from (\ref{unbiasedness}). Based on the above discussion we assume that 
\begin{equation}
\bar{\boldsymbol{N}}=0\qquad\textrm{or}\qquad\left\langle \boldsymbol{N}\right\rangle =0\quad\textrm{and}\quad\forall_{l}\left\langle \partial_{l}S\cdot N_{l}\right\rangle =0.
\end{equation}
This unified condition allows that the momentum fluctuations either vanish on average in the strong sense, or only in the weak sense being componentwise unbiased with the classical momentum.

\subsection{The exact uncertainty principle}

Still following \cite{Hall}, we denote by $\delta x$ a ``direct measure of uncertainty in position''. The EUP states that $\delta x$ is fully characterized by $P$ and, most importantly, that the total fluctuations are inversely correlated with $\delta x$. In other words, if one considers a rescaling transformation $P\left(x\right)\mapsto P_{\kappa}\left(x\right)=\kappa^{n}P\left(\kappa x\right),$ then since $P_{\kappa}\left(x\right)$ is narrower (broader) for $\kappa>1$ ($\kappa<1$ ), the position uncertainty accordingly transformed as $\delta x\mapsto\kappa^{-1}\delta x$ becomes smaller (bigger) in comparison with the initial one. The inverse law of the EUP implies the linear transformation of the total momentum fluctuations 
\begin{equation}
\Delta N\mapsto\kappa\Delta N.\label{trMom}
\end{equation}

Further analysis of the above scaling relation \cite{Hall} leads to the solution for $(\Delta N)^{2}$ of the form:
\begin{equation}
(\Delta N)^{2}=C\sum_{l=1}^{n}F_{l}\left[P\right],\qquad C=\hbar^{2}/4.\label{przez Fishery}
\end{equation}
Note that in the case of independent subsystems, the particular fluctuations associated with every direction shall also be mutually independent. This property together with (\ref{deltaN}) and (\ref{przez Fishery}) uniquely fix $\Delta N_{l}=\sqrt{CF_{l}\left[P\right]}$. It is also important to emphasize here the spherical symmetry related to the rescaling transformation. In fact, not only the total fluctuations become multiplied by $\kappa$, but the same scaling property independently applies to every $\Delta N_{l}$. As a result, $(\Delta N)^{2}$ is invariant with respect to rotations of the coordinate space, and as a functional can only be made with invariant quantities such as $\boldsymbol{\nabla}P\cdot\boldsymbol{\nabla}P$. This property has been implicitly used in \cite{Hall} as the starting point in the derivation of (\ref{przez Fishery}). 

Finally, if one \emph{decides} to describe the position uncertainty in the $l$th direction by the square root of the inverse of the Fisher information, $\delta x_{l}=1/\sqrt{F_{l}\left[P\right]}$, one finds the equality of the form
\begin{equation}
\delta x_{l}\Delta N_{l}=\frac{\hbar}{2}.\label{EUP}
\end{equation}
Note that $\delta x_{l}$ is a good, though not required, choice for the direct uncertainty measure $\delta x$ in the 1D case. From the well known Cramer-Rao bound one has $\Delta x_{l}\geq\delta x_{l}$ while, as explained in Sec. \ref{sec:Fluct}, $\Delta p_{l}\geq\Delta N_{l}$. As a corollary from the EUP one thus obtains the HUR, satisfied independently for all $l\in\left\{ 1,\ldots,n\right\} $.

The Euler\textendash Lagrange equations for the quantum action $\mathcal{A}_{Q}$ reproduce the continuity equation (\ref{continuity}) and lead to the modified Hamilton-Jacobi equation of the form
\begin{equation}
\frac{\partial S}{\partial t}+\frac{1}{2m}\boldsymbol{\nabla}S\cdot\boldsymbol{\nabla}S+\frac{\hbar^{2}}{8m}\left[\frac{1}{P^{2}}\boldsymbol{\nabla}P\cdot\boldsymbol{\nabla}P-\frac{2}{P}\nabla^{2}P\right]+V=0.\label{newJacobi}
\end{equation}
Finally, by a subtle substitution
\begin{equation}
S=-i\hbar\ln\frac{\psi}{\sqrt{P}},\quad\textrm{equivalently given as }\quad\psi=\sqrt{P}e^{iS/\hbar},\label{Podstawienie}
\end{equation}
one transforms (\ref{newJacobi}) to be the pure Schr{\"o}dinger equation: 
\begin{equation}
-i\hbar\frac{1}{\psi}\frac{\partial\psi}{\partial t}-\frac{\hbar^{2}}{2m}\frac{1}{\psi}\nabla^{2}\psi+V=0.
\end{equation}
From (\ref{Podstawienie}) one also gets that $P=\left|\psi\right|^{2}\equiv\rho$.

\section{Generalized exact uncertainty principle \label{sec:Generalized-Exact-Uncertainty}}

Having in mind the GUP, Eq. (\ref{GUP}), as a model generalization of the HUR, we would like to study its possible interrelations with the EUP. First of all, if a small value of the scaling parameter $\kappa$ is considered, the momentum fluctuations remain small and the term $(\Delta p)^{2}$ present on the right hand side of (\ref{GUP}) does not play a significant role. For bigger values of $\kappa$, however, as the momentum fluctuations are assumed to grow linearly with $\kappa$, the more visible right hand side contribution might violate the GUP. Since the term proportional to $(\Delta p)^{2}$ is multiplied by a very small parameter, in the regime in which it actually makes a physical sense to discuss (\ref{GUP}), the quadratic contribution shall remain majorized by the basic linear term from the left hand side. To make this mechanism work it will be desired to let the momentum fluctuations grow superlinearly. The overall message is thus that if generalized (sharper) forms of the quantum uncertainty relation are expected instead of the HUR, the EUP shall deviate from the law of inverse correlation.

To follow the intuition sketched above, let us modify the EUP scaling law (\ref{trMom}) to the form
\begin{equation}
\Delta N_{l}\mapsto w^{-1}\left(\kappa w\left(\Delta N_{l}\right)\right),\label{GenEUP}
\end{equation}
or 
\begin{equation}
w\left(\Delta N_{l}\right)\mapsto\kappa w\left(\Delta N_{l}\right),
\end{equation}
with $\kappa$ presented as the proper scaling factor. The difference with respect to (\ref{trMom}) is due to a non-negative, increasing function $w\left(\cdot\right)$ such that:
\begin{equation}
\lim_{z\rightarrow0}w\left(z\right)=0,\qquad\lim_{z\rightarrow0}w'\left(z\right)=1,\label{conditions}
\end{equation}
and (most importantly) 
\begin{equation}
w\left(z\right)<z,\qquad\textrm{whenever}\qquad z>0.\label{lastReq}
\end{equation}
The first requirement from (\ref{conditions}) assures that the new scaling transformation does not generate fluctuations if they are initially absent. The second condition states that for infinitesimal values of $\Delta N_{l}$ the formula (\ref{GenEUP}) does not differ from the original EUP scaling. The last requirement (\ref{lastReq}) is the crucial ingredient of the EUP modification which simply aims to build up the fluctuations whenever $\kappa>1$. As an easy example, if $w\left(z\right)=z^{1/s}$, with $s>1$, the modified scaling law would give $\Delta N_{l}\mapsto\kappa^{s}\Delta N_{l}$. The nested structure ($w^{-1}$ on top) is necessary since for $\kappa=1$ the transformation shall be the identity.

We start the discussion of the modified EUP (\ref{GenEUP}) with an observation, that the functions $w$ different from identity break the spherical symmetry of $(\Delta N)^{2}$. Note however, that if we simply redefine the fluctuation components as $\Delta N_{l}^{(w)}=w\left(\Delta N_{l}\right)$, we recover the whole structure of the original EUP. The quantity $\Delta N^{(w)}$, defined via the formula (\ref{deltaN}) with $\Delta N_{l}$ replaced by $\Delta N_{l}^{(w)}$, not only is inversely correlated with $\delta x$, but also enjoys the desired spherical symmetry. The derivation of the formula (\ref{przez Fishery}) presented in \cite{Hall} can thus immediately be repeated for $\left(\Delta N^{(w)}\right)^{2}$. Moreover, the function $w^{-1}$ preserves the argumentation based on the independent subsystems, so that $\Delta N_{l}^{(w)}=\sqrt{CF_{l}\left[P\right]}$. The last observation is valid since in order to make $\Delta N_{l}$ independent (for separable case), also $\Delta N_{l}^{(w)}$ need (this is also sufficient) to be independent.

As a counterpart of Eq. \ref{EUP} we obtain
\begin{equation}
\delta x_{l}\Delta N_{l}^{(w)}=\frac{\hbar}{2}=\delta x_{l}w\left(\Delta N_{l}\right).\label{yetNewEUP}
\end{equation}
Since the function $w$ is increasing, the relation $\Delta p_{l}\geq\Delta N_{l}$ implies $w\left(\Delta p_{l}\right)\geq w\left(\Delta N_{l}\right)$. Directly from (\ref{yetNewEUP}) we obtain the following, modified uncertainty relation
\begin{equation}
\Delta x_{l}w\left(\Delta p_{l}\right)\geq\frac{\hbar}{2},\label{SharperHUR}
\end{equation}
valid, for any $l$, instead of the traditional HUR. Note that due to (\ref{lastReq}), this new form of UR is sharper than the HUR. 

For instance, if we want (\ref{SharperHUR}) to be equal to the GUP, we need to set
\begin{equation}
w\left(z\right)=\frac{z}{1+\beta z^{2}}.
\end{equation}
The above function is increasing as long as $z\leq\beta^{-1/2}$, so that the range $0\leq\Delta p\leq\beta^{-1/2}$ needs to be taken as the domain of validity for our approach in the GUP case. Assuming $\beta_{0}=1$, this gives a restriction of order $\Delta p\leq\hbar/l_{p}$, so the upper bound for $\Delta p$ approximately equal to $10^{23}m_{e}c$, with $m_{e}$ being a mass of an electron. Actually, the GUP contribution to HUR can be treated as a correction only under this limitation (otherwise it becomes a dominant term and higher-order corrections need to be taken into account). Only if $\Delta p\leq\beta^{-1/2}$, the superlinear scaling of the momentum fluctuations encompasses the $\beta\left(\Delta p\right)^{2}$ contribution in (\ref{GUP}).

The new form of the fluctuation functional, $\Delta N_{l}=w^{-1}\left(\sqrt{CF_{l}\left[P\right]}\right)$, after being inserted into the action $\mathcal{A}_{Q}$, modifies the quantum Hamilton-Jacobi equation (\ref{newJacobi}) to the form
\begin{equation}
\frac{\partial S}{\partial t}+\frac{1}{2m}\boldsymbol{\nabla}S\cdot\boldsymbol{\nabla}S+\frac{\hbar^{2}}{8m}\sum_{l=1}^{n}\left(1+W_{l}\left[P\right]\right)\left[\frac{1}{P^{2}}\left(\frac{\partial P}{\partial x_{l}}\right)^{2}-\frac{2}{P}\frac{\partial^{2}P}{\partial x_{l}^{2}}\right]+V=0,\label{modifiedJacobi}
\end{equation}
with 
\begin{equation}
W_{l}\left[P\right]=W\left(CF_{l}\left[P\right]\right),
\end{equation}
and the function $W\left(\cdot\right)$ equal to
\begin{equation}
W\left(z\right)=\frac{d}{dz}\left[w^{-1}\left(\sqrt{z}\right)\right]^{2}-1.\label{W}
\end{equation}
The continuity equation (\ref{continuity}) once more stays untouched. In our model example of the GUP we easily find that
\begin{equation}
W\left(z\right)=\frac{2}{1+\sqrt{1-4\beta z}-2\beta z\left(2+\sqrt{1-4\beta z}\right)}-1=4\beta z+\mathcal{O}(\beta^{2}).\label{GUPfunction}
\end{equation}

The terms proportional to $W$ are naturally responsible for the nonlinear corrections to the Schr{\"o}dinger equation. The substitution (\ref{Podstawienie}), $P=\rho$, and the identity
\begin{equation}
\frac{1}{\rho^{2}}\left(\frac{\partial\rho}{\partial x_{l}}\right)^{2}-\frac{2}{\rho}\frac{\partial^{2}\rho}{\partial x_{l}^{2}}=-\frac{4}{\left|\psi\right|}\frac{\partial^{2}\left|\psi\right|}{\partial x_{l}^{2}},
\end{equation}
bring Eq. \ref{modifiedJacobi} to the final form (\ref{NLSchr}).

\section{Harmonic oscillator example\label{sec:Harmonic-oscillator-example} }

We would now briefly like to study eventual manifestations of the modified quantum stationary dynamics. As mentioned in the Introduction the free-particle case, $V=0$, is trivial as for the plane wave solutions $\left|\psi\right|$ is a constant. This fact implicitly implies that the plane waves remain the quantum states with $\Delta p=0$, even though the whole EUP formalism does not look for the operator definition of momentum.

Let us thus work out the one-dimensional, stationary case of the harmonic potential $V=\frac{1}{2}\zeta x^{2}$:
\begin{equation}
E\psi=\left(-\frac{\hbar^{2}(1+\nu)}{2m}\frac{d^{2}}{dx^{2}}+\frac{1}{2}\zeta x^{2}\right)\psi,
\end{equation}
with $\nu=W\left(CF\left[\rho\right]\right)$. Since in the textbook solutions for the harmonic oscillator, all the energy eigenstates possess real wave functions (this fact is already included in the above equation), there is no obstacle in performing the complete analysis of this problem. For simplicity, we restrict here however only to the (normalized) ground state given by:
\begin{equation}
\psi_{0}(x;\nu)=\left(\frac{1}{\pi\sigma^{2}}\right)^{1/4}e^{-x^{2}/2\sigma^{2}},\qquad\sigma^{2}=\sigma_{0}^{2}\sqrt{1+\nu},\qquad\sigma_{0}^{2}=\sqrt{\frac{\hbar^{2}}{\zeta m}}.
\end{equation}
The Fisher information of this state is
\begin{equation}
F[\rho_{0}]=\frac{2}{\sigma^{2}}\equiv\frac{2}{\sigma_{0}^{2}\sqrt{1+\nu}}.
\end{equation}

\begin{figure}
\begin{centering}
\includegraphics[scale=0.6]{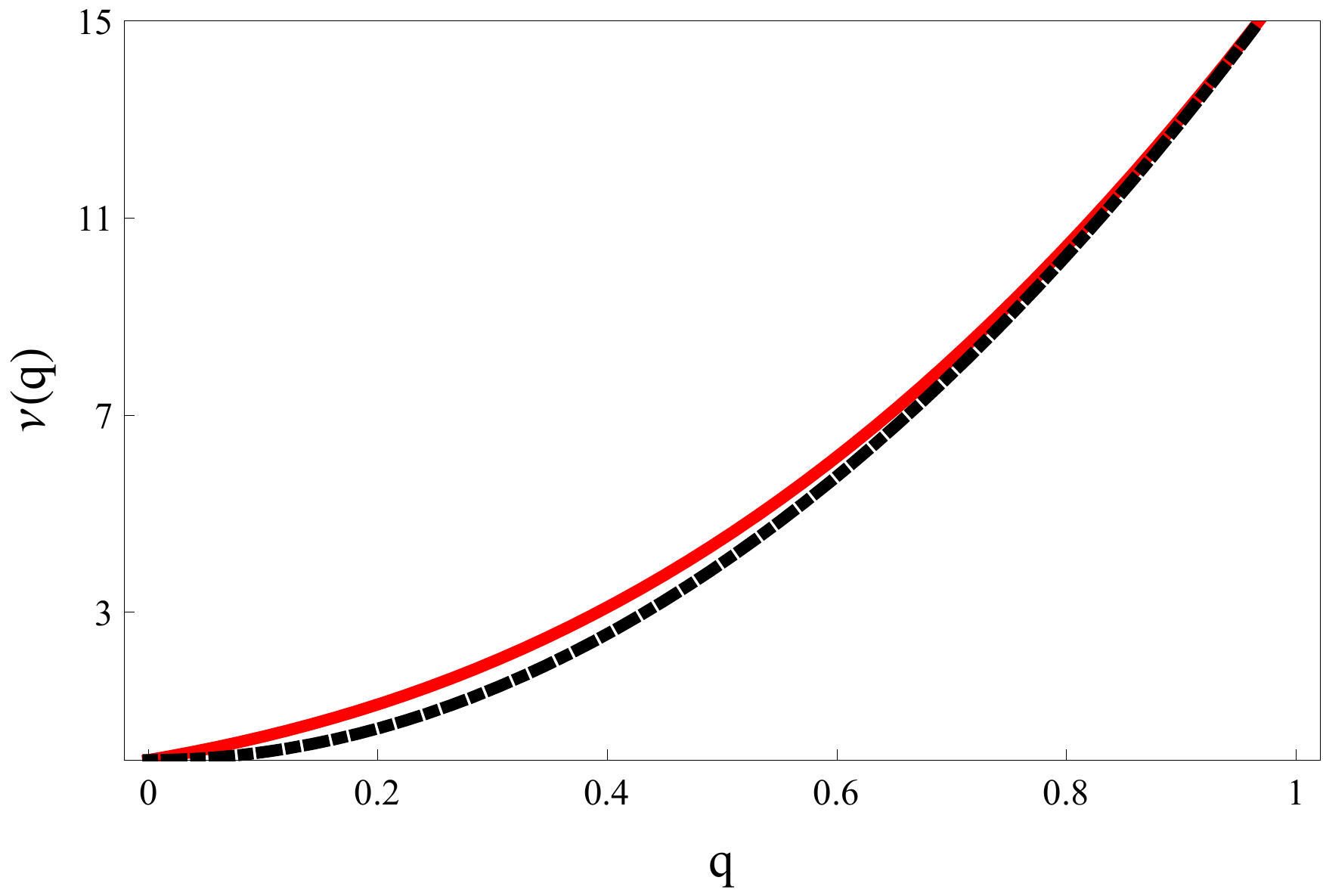}
\par\end{centering}

\protect\caption{\label{rys} The function $\nu\left(q\right)$ (red, solid) together with its asymptotic form $16q^{2}$ (black, dotted). }

\end{figure}
While focusing on the GUP example, we need to use the function (\ref{GUPfunction}), to determine the value of $\nu$. We find:
\begin{equation}
\nu\left(q\right)=\frac{q}{1+q^{2}}\left(4\sqrt{1+q^{2}}+q\left[7+8q\left(q+\sqrt{1+q^{2}}\right)\right]\right),\label{vq}
\end{equation}
where $q=\hbar^{2}\beta/(2\sigma_{0}^{2})$. This potentially nontrivial function actually very quickly (as can be seen on Fig. \ref{rys}) becomes indistinguishable from $16q^{2}$, which represents the asymptotic behavior of $\nu\left(q\right)$ for large $q$. We shall use this fact to anticipate that $\lim_{q\rightarrow\infty}q^{-1}\sqrt{1+\nu\left(q\right)}=4$.

Obviously for larger values of the basic variance $\sigma_{0}^{2}$, the effect of the above modification is negligible. The theory, however nicely predicts the minimal position uncertainty which for the ground state of the harmonic oscillator is given by:
\begin{equation}
\min\left(\Delta x\right)^{2}=\frac{1}{2}\min_{\sigma_{0}\geq0}\sigma^{2}=\frac{1}{2}\min_{\sigma_{0}\geq0}\sigma_{0}^{2}\sqrt{1+\nu\left(\hbar^{2}\beta/(2\sigma_{0}^{2})\right)}.
\end{equation}
If we change the variable to $q$ 
\begin{equation}
\min\left(\Delta x\right)^{2}=\frac{\hbar^{2}\beta}{4}\min_{q\geq0}\frac{1}{q}\sqrt{1+\nu\left(q\right)},
\end{equation}
and calculate the derivative
\begin{equation}
\frac{d}{dq}\left(\frac{1}{q}\sqrt{1+\nu\left(q\right)}\right)=-\frac{\sqrt{1+q^{2}}+2q\left[1+q\left(q+\sqrt{1+q^{2}}\right)\right]}{q^{2}\left(1+q^{2}\right)^{2}\sqrt{8q^{2}\left(1+q^{2}\right)+1+4q\sqrt{1+q^{2}}\left(1+2q^{2}\right)}}
\end{equation}
which happens to be always negative (so that the function to be minimized is decreasing), we can conclude that
\begin{equation}
\min\left(\Delta x\right)^{2}=\frac{\hbar^{2}\beta}{4}\lim_{q\rightarrow\infty}q^{-1}\sqrt{1+\nu\left(q\right)}=\hbar^{2}\beta\equiv\beta_{0}l_{p}^{2}.
\end{equation}
As already mentioned, obtaining the limit is straightforward since for large $q$ one finds $\nu\left(q\right)\sim16q^{2}$ as only the very last terms from (\ref{vq}) contribute. In the next section we explain why the above observation is universally valid for any solution of the discussed equation.

Exactly the same result, could be obtained by optimizing the GUP with respect to $\Delta p$. This known fact follows from 
\begin{equation}
\min\Delta x=\hbar\min_{\Delta p\geq0}\left[\frac{1+\beta\left(\Delta p\right)^{2}}{2\Delta p}\right]=\hbar\sqrt{\beta},
\end{equation}
which is a basic optimization problem.

\section{Conclusions}

The main achievement of this contribution is the new proposal for the nonlinear Schr{\"o}dinger equation which enjoys a number of desired properties, namely, separability for non-interacting subsystems, homogeneity and Galilean invariance. It can thus serve as a support for future understanding of the quantum evolution origins and the validity of its linearity. Moreover, this equation has been derived from the generalized form of the exact uncertainty principle, which serves as a prerequisite for the generalized versions of the Heisenberg uncertainty relation, such as the GUP (\ref{GUP}).

Using the GUP as a toy model introducing additional gravitational effects to the standard description of quantum uncertainty, we shall point out two important observations. First of all, while thinking about the relevance of the presented nonlinearities (in particular, in the GUP context) one could (and actually should) question the marriage of classical mechanics and extreme quantum regimes necessarily related to the nonlinearities. Even though, on a first impression this match seems to be conceptually problematic, it happens to be able to capture fundamental features of quantum systems under discussion, such as the minimal uncertainty of the position variable. In Sec. \ref{sec:Harmonic-oscillator-example} this property is shown for the ground state of the harmonic oscillator. It is however of general validity since the function $W(z)$ in (\ref{GUPfunction}) is singular for $z=0$ and $z=1/\left(4\beta\right)$, while it becomes complex whenever $z>1/\left(4\beta\right)$. The first case is not relevant because $z=CF\left[\rho\right]$ is equal to $0$ only for plane waves, for which the derivatives of $\left|\psi\right|$ do vanish anyway. Due to the two remaining issues we necessarily have $F\left[\rho\right]\leq1/\left(\hbar^{2}\beta\right)$, and the physical system needs to acquire infinite energy to saturate the inequality. This last result, by virtue of the Cramer-Rao bound is equivalent to existence of the minimal observable distance equal to $\hbar\sqrt{\beta}$.

The second observation is the main physical insight of the approach --- the fact that plane waves do not feel the nonlinear interaction. Since it is possible to extend the basic EUP formalism to the case of the Klein Gordon equation \cite{Goss} there is no obstacle to directly apply the results of this paper in the relativistic case. Without going into the details (an extended analysis could be performed in the future) I conclude that gravitational effects added to the standard relativistic dynamics do not need to affect plane-wave solutions. It does not need to be true (even though it is commonly believed \cite{Plato,Montevideo}) that dispersion relations of the plane waves are affected on the Planck scale thus carrying a signature of quantum gravitational effects. Moreover, the issues with Lorentz invariance could potentially be solved in the same way (relativistic dynamics together with generalized EUP) if one starts from the relativistic counterpart of the GUP. The last requirement is conceptually a bit difficult. The proposal related to the maximal proper acceleration approach \cite{GUPpA} seems to be promising in that context.

Staying with the GUP model I would like to emphasize another particularly valuable content of the formalism presented. In \cite{GUPPRD}, the Schr{\"o}dinger equation in the GUP case has been derived in the momentum representation. Issues related to the description of the position-momentum duality made it impossible to develop the equation in the position domain. Moreover, the momentum representation equation, while linear, contained momentum derivatives of every order. On the contrary, Eq. \ref{NLSchr} stays on the ground of the position representation, and due to its classical roots actually knows nothing about the operator nature of both position and momentum variables. This new evolution equation, while quasi-nonlinear, involves only second order spatial derivatives. It can be thus treated as the opposite side of the mirror, in comparison with the GUP Schr{\"o}dinger equation from \cite{GUPPRD}. 

Last, but not least, there is a single contribution \cite{parwani} which utilizes the EUP approach to discuss nonlinearities of the Schr{\"o}dinger dynamics. The resulting equation differs substantially from the one derived here as it again contains derivatives of any order. It could be interesting to have a closer look at interrelations between the two approaches, also in the context of plane waves.

\ack{}{}

It is my pleasure to thank Michael Hall and Marcel Reginatto as well as Gaetano Lambiase for interesting and helpful correspondence. Financial support from Grant No. 2014/13/D/ST2/01886 of the Polish National Science Centre is gratefully acknowledged. Research in Cologne is supported by the Excellence Initiative of the German Federal and State Governments (Grant ZUK 81) and the DFG (GRO 4334/2-1). I acknowledge hospitality of Freiburg Center for Data Analysis and Modeling.

\end{document}